\documentstyle[amssymb,aps,prl,psfig]{revtex}

\begin{document}
\twocolumn[\hsize\textwidth\columnwidth\hsize\csname
@twocolumnfalse\endcsname

\title{Three-wave mixing of Bogoliubov quasi-particles in a Bose condensate }
\author{R. Ozeri,  N. Katz, J. Steinhauer and N. Davidson}
\address{Department of Physics of Complex Systems,\\
Weizmann Institute of Science, Rehovot 76100, Israel}

\maketitle

\begin{abstract}
A dressed basis is used to calculate the dynamics of three-wave
mixing between Bogoliubov quasi-particles in a Bose condensate.
Due to the observed oscillations between different momenta modes,
an energy splitting, analogous to the optical Mollow triplet,
appears in the Beliaev damping spectrum of the excitations from
the oscillating modes.
\end{abstract}

 ]

Since the experimental realization of Bose-Einstein condensation
in trapped atomic gases, which serves as a mono-energetic and
dense atomic source, experiments in non-linear atom optics have
become feasible. Atomic four-wave mixing (4WM) \cite{Phillips
fwm}, Superradiance \cite{Superadiance exp}, and matter-wave
amplification \cite{Amplification Kett}, \cite{Amplification
Japan} are all examples of non-linear atom optics, which involve
the mixing of several atomic fields or the mixing of atomic and
electromagnetic fields.

In the case of atomic 4WM, bosonic amplification directs one of
the products of a collision between an excitation atom and a
condensate atom into an, initially, largely populated mode
\cite{Phillips fwm},\cite{Julienne1},\cite{Myestre1}. Thus, a new
mode is macroscopically populated through the second collision
product. When the energy, $\varepsilon _{k}$, of an excitation
mode $\mathbf{k}$ is low compared to the condensate chemical
potential, the excitation can no longer be described as a free
atom moving with momentum $\mathbf{k}$, but rather as a collective
phonon excitation, which involves a large number of
atoms\cite{Vogels}. Atomic 4WM is therefore inadequate for the
description of phonon decay. Phonon excitations are described in
the framework of Bogoliubov theory \cite{Bogoliubov}. As
Bogoliubov excitations are bosons, bosonic amplification will also
direct one of the damping products of a phonon into an, initially,
largely populated mode, leading to three wave-mixing (3WM) of
Bogoliubov quasi-particles.

In this letter we introduce a basis of states which are dressed by
the interaction between three, largely populated, modes of
Bogoliubov quasi-particles in a Bose-Einstein condensate at zero
temperature. Using this basis of quantum states we calculate the
time evolution of the system, which exhibits non-linear
oscillations between the different momentum modes. The use of the
wave-mixing eigen-basis enables the calculation of the system
dynamics efficiently and with no approximations. In contrast to
other theoretical approaches to wave-mixing, which lead to
non-linear differential equations \cite{Julienne1},\cite{Burnett1}
the dressed state approach turns the problem into a linear one,
for which propagation in time is trivial. We show that, due to
relative number squeezing, the variance in the number difference
between the two low momentum modes remains constant.

In analogy to the treatment of spontaneous photon scattering in
the atom-laser dressed system \cite{CCT}, we treat damping into
the quasi-continuum of empty modes, during the 3WM dynamics, as
transfer between dressed state manifolds. Thus, the effects of 3WM
on the damping process, are calculated. The damping energy
spectrum is presented. A transition from elastic to inelastic
damping is observed, which, in analogy to the optical Mollow
triplet, leads to a splitting of the spectrum into a doublet of
resonance energies.

Previously, nonlinear mixing of quasi-particles in a Bose
condensate was studied, between the two lowest energy excitation
modes in the discrete regime, where the energy of the excited mode
is of the order of the energy separation between modes
\cite{Burnett1}, \cite{Burnett2}, \cite{Foot1}.

Our model system is a homogenous condensate of finite volume $V$,
with $ N_{0}$ atoms in the ground state. The Hamiltonian for the
system in the Bogoliubov basis, taken to the $\sqrt{N_0}$ order,
is given by  \cite{Fetter}

\begin{equation}
H=H_{0}+H_{int},
\end{equation}

where $H_{0}=\frac{1}{2}gnN_{0}-\sum_{\mathbf{k}}\varepsilon
_{k}v_{k}^{2}+\sum_{\mathbf{k}}\varepsilon
_{k}b_{\mathbf{k}}^{+}b_{\mathbf{k}}$ is the part of $H$ which is
diagonalized by the Bogoliubov transformation. $N_0$ is the number
of atoms in the $\mathbf{k}=0$ mode. $\varepsilon
_{k}=\sqrt{\varepsilon _{k}^{0}\left( \varepsilon
_{k}^{0}+2gn\right) }$ is the recently measured Bogoliubov energy
\cite{our experiment1} and $\varepsilon _{k}^{0}$ is the free
particle energy $\varepsilon _{k}^{0}=\hbar ^{2}k^{2}/2m$. $g=4\pi
\hbar ^{2}a/m$ is the coupling constant, $a$ is the s-wave
scattering length, $m$ is the mass of the atoms.
$b_{\mathbf{k}}^{+}$ and $b_{\mathbf{k}}$ are the creation and
annihilation operators respectively, of quasi-particle excitations
with momentum $\mathbf{k}$. $v_k$ and $u_k$ are the Bogoliubov
quasi-particle amplitudes. This part of the Hamiltonian describes
excitations in the condensate with momentum $\mathbf{k}$ and
energy $\varepsilon _{k}$.

$H_{int}$ is the part of the Hamiltonian which is responsible for
the interaction between excitations

\begin{equation}
H_{int}=\frac{g}{2V}\sqrt{N_{0}}\sum_{\mathbf{k},\mathbf{q}}
A_{kq}\left(
b_{\mathbf{k}}^{+}b_{\mathbf{q}}b_{\mathbf{k}-\mathbf{q}}+b_{\mathbf{q}}^{+}b_{\mathbf{k}-\mathbf{q}}^{+}b_{\mathbf{k}}\right).
\end{equation}

The first term in parentheses is referred to as Landau damping and
is analogous to photonic up conversion, such as second harmonic
generation in optics \cite{Pitaevskii}, whereas the second term is
referred to as Beliaev damping, and is analogous to photonic down
conversion \cite{Giorgini}. $A_{kq}$ is the many-body suppression
factor

\begin{eqnarray}
&& A_{kq}=2u_{k}\left(
u_{q}u_{k-q}-v_{q}u_{k-q}-u_{q}v_{k-q}\right)\\ \nonumber
&&-2v_{k}\left( v_{q}v_{k-q}-u_{q}v_{k-q}-v_{q}u_{k-q}\right).
\end{eqnarray}

The main damping mechanism from a single, largely populated mode
$\mathbf{k}$, will be elastic Beliaev damping of the excitations
into empty modes, which are on a mono-energetic surface in
momentum space \cite{our experiment3}.

We now consider the case where the condensate is excited with $N$
excitations of momentum $\mathbf{k}$ and $M$ excitations of
momentum $\mathbf{q}$, $\left| N _{\mathbf{k}},
M_{\mathbf{q}},0_{\mathbf{k}-\mathbf{q}}\right\rangle $, such that
$\mathbf{k}$ and $\mathbf{q}$ fulfill the Bragg condition, i.e.
$\varepsilon _{\mathbf{k}}=\varepsilon _{\mathbf{q}}+\varepsilon
_{\mathbf{k}-\mathbf{q}}$. Two-photon Bragg transitions can be
used to excite the condensate, and populate different momentum
modes with a variable number of excitations \cite{our
experiment1}. The amplitude given by $H_{int}$ for Beliaev damping
of the $\mathbf{k}$ momentum excitation into two excitations with
momenta $\mathbf{q}$ and $\mathbf{k}-\mathbf{q}$, $\left| \left(
N-1\right) _{\mathbf{k}},\left( M+1\right)
_{\mathbf{q}},1_{\mathbf{k}-\mathbf{q}}\right\rangle $, will be
$\sqrt{N}$ or $\sqrt{M}$ fold larger than any of the other damping
channels. This will result in 3WM into a newly populated
$\mathbf{k}-\mathbf{q}$ momentum mode. We assume that
$N_0>>N,M>>\Gamma{t_0}$, where $\Gamma$ is the total Beliaev
damping rate of the excitations into the quasi-continuum of empty
modes and $t_0$ is the time of the experiment. Thus, during the
experiment time, 3WM dynamics dominates over damping into empty
modes. In general the set of $N+1$ excitation Fock states $\left|
\left( N-i\right) _{\mathbf{k}},\left( M+i\right)
_{\mathbf{q}},i_{\mathbf{k}-\mathbf{q}}\right\rangle $, where $i$
varies between $0$ and $N$, spans a degenerate subspace of the
eigenstates of $H_{0}$. $H_{int}$ couples between pairs of states
in this subspace which have a difference of $1$ in $i$, and can be
represented by the $(N+1)\times{(M+1)}$ tri-diagonal matrix

\begin{eqnarray}
\left\langle N-s,M+s,s\right|H_{int}\left|N-i,M+i,i\right\rangle = \nonumber \\
\frac{g}{2V}\sqrt{N_{0}}A_{kq} ( \sqrt{N-i}\sqrt{M+i+1}\sqrt{i+1}\delta _{i+1,s} + \nonumber \\
\sqrt{N-i+1}\sqrt{M+i}\sqrt{i}\delta _{i-1,s} ).
\end{eqnarray}

When $H_{int}$ is diagonalized, we get a new set of $N+1$
eigenstates, $\left| j\right\rangle $, where $j$ varies between
$1$ and $N+1$, that are dressed by the interaction. Note that
since the dressed states are superpositions of degenerate
eigenstates of $H_{0}$, they are eigenstates of the complete
Hamiltonian in Eq. (1).

The filled circles in Fig. 1 show the calculated energy spectrum of the dressed basis for $%
N=M=100$ in units of $(g\sqrt{N_0}/{2V})A_{kq}$. The degeneracy
between the excitation Fock states is removed by the interaction.
The spectrum appears to be roughly linear with an average energy
spacing of ~$dE\simeq 2.48\sqrt{N}$\, in the above units
\cite{scaling law}. The hollow circles in Fig. 1 show the value of
the energy difference between each two dressed states. Due to the
non-linearity of the problem, energy differences between each pair
of dressed states are slightly different.

\begin{figure}[h]
\begin{center}
\mbox{\psfig{figure=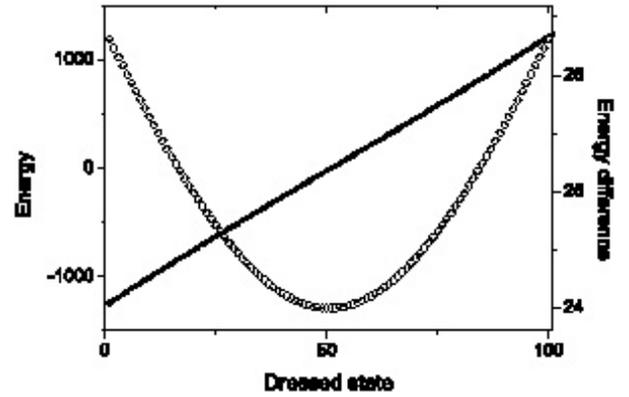,width=8.0cm}}
\end{center}
\vspace{0.4cm} \caption{Filled circles show the spectrum of the
$M=N=100$ manifold in units of ${g\sqrt{N_0}\over{2V}}A_{kq} $.
Hollow circles show the energy difference between each two dressed
states in the same units. The spectrum is not linear, and the
energy spacing varies parabolically around an average of
$2.48\sqrt{N}$.}
\end{figure}

\begin{figure}[h]
\begin{center}
\mbox{\psfig{figure=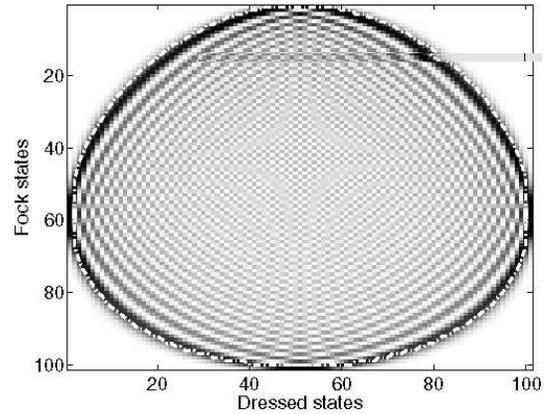,width=7.0cm}}
\end{center}
\vspace{0.4cm} \caption{The square of the transfer matrix between
the excitation Fock state basis and the dressed state basis for
$N=M=100$. Darker areas correspond to larger probability. The
white dashed-dotted line draws the solution to
$\left|E_j\right|=2\sqrt{(N^2-i^2)i}$. The Fock states are
numbered by $i$. The dressed states are numbered by $j$ from
lowest to highest energy.}
\end{figure}

Figure 2 shows the absolute value squared of the transfer matrix
between the bare excitation Fock state basis and the dressed basis
for $N=M=100$. Since $H_{int}$, after being applied a sufficient
number of times, can couple each two bare states, most dressed
states are spanned by a large number of bare states \cite{Fock}.

In the limit of $M=N$ $\rightarrow \infty $ we can use the
tri-diagonality of $H_{int}$ to approximate the dressed state as
the solution to the differential equation

\begin{equation}
\frac{\partial a_{j}^{2}\left( x\right) }{\partial x^{2}}=\left( \frac{%
\lambda _{j}}{\sqrt{\left( 1 - x^{2}\right)x}}-{f_0}^2\right)
a_{j}\left( x\right),
\end{equation}

with the boundary conditions ${da_{j}\left( x\right)}/{dx} =\left(
\lambda _{j}/\sqrt{N}-N\right) a_{j}\left( x\right)$ at $x=0$, and
${da_{j}\left( x\right)}/{dx}=\left( { N- \lambda
_{j}/\sqrt{2N}}\right) a_{j}\left( x\right) $ at $x=1$. Here
$dx=1/N$, $x=idx$, and $a_{j}\left( x\right) =\left\langle
N-i,M+i,i | j\right\rangle $ is the projection of $\left|
j\right\rangle $ on the Fock state with $i$ excitations removed
from the $\mathbf{k}$ momentum mode. Also, $\lambda
_{j}=E_{j}\sqrt{N}$, $E_{j}$ is the energy eigenvalue of $\left|
j\right\rangle$, and ${f_0}^2=2{N}^2$. The main contribution to
the dressed state superposition of Fock states comes from the two
states which solve $\left( \frac{\left|\lambda
_{j}\right|}{\sqrt{\left( 1 - x^{2}\right)x}}-{f_0}^2\right)=0$.
The dashed-dotted line in Fig. 2 draws the solution of this
equation. In order for this equation to have real solutions the
resulting spectrum must satisfy $\left|E_{j}\right|< 1.24N^{3/2}$.
Since there are $N+1$ dressed states we get an average energy
difference between dressed states of $dE\simeq 2.48\sqrt{N}$,
consistent with our numerical observation.

\begin{figure}[h]
\begin{center}
\mbox{\psfig{figure=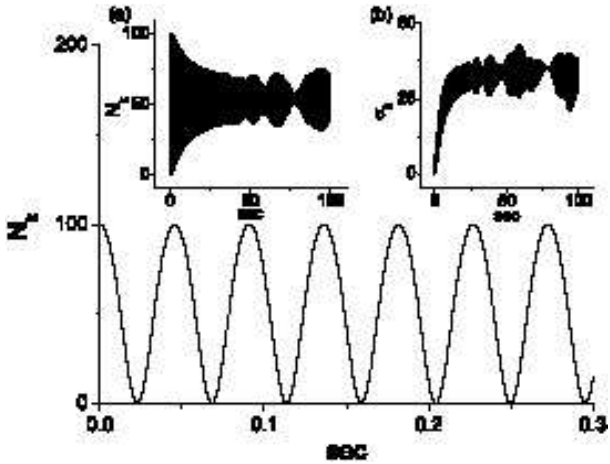,width=8.0cm}}
\end{center}
\vspace{0.4cm} \caption{The expectation value $N_{\mathbf{k}}$,
$k=0.7$ $\hbar/\xi$ and $q=k/\sqrt2$, as a function of time.
$N=M=100$, for a condensate of $3\times{10^5}$ $^{\text{87}}$Rb
atoms in the $F=2$, $mf=2$ ground state. The condensate is
homogeneous, with a density of $3\times{10^{14}}$ $atoms/cm^3$.
The oscillation frequency is roughly the average energy difference
in the dressed state spectrum. Insets (a) and (b) show
$N_{\mathbf{k}}$ and the standard deviation of $N_{\mathbf{k}}$
respectively, during a longer time.}
\end{figure}

We choose as a model system a condensate of $3\times10^{5\text{
}87}Rb$ atoms in the $F=2$, $m_{f}=2$ ground state. The
condensate, which is similar to the experimental parameters of
\cite{our experiment1} is homogeneous and has a density of
$3\times10^{14}$ $atoms/cm^{3}$. The $k=0.7$ $\hbar/\xi$ mode and
$q=k/\sqrt2$ mode, where $\xi $ is the healing length of the condensate given by $\xi =%
\sqrt{8\pi na}^{-1}$, are populated with a 100 excitations each.

We start from the above excitation Fock state, written as a linear
superposition of dressed states, which are obtained by
diagonalizing (4). The state of the system is then readily
propagated in time by evolving each of the dressed states phases
according to its energy. Figure 3 shows $N_{\mathbf{k}}$, the
expectation value of the number of excitations with momentum
$\mathbf{k}$, as a function of time. Excitations oscillate between
the $\mathbf{k}$ momentum mode and the $\mathbf{q}$ and
$\mathbf{k}-\mathbf{q}$ momenta modes. Inset (a) of Fig. 3 shows
$N_k$ during a much longer time. Since the energy spectrum is not
precisely linear, at longer times beating between the different
oscillation frequencies gives rise to a slow amplitude modulation
of the oscillations.

Even though we start from an excitation Fock state, the system
immediately evolves into a superposition of Fock states. Inset (b)
of Fig. 3 shows the standard deviation of $N_{\mathbf{k}}$ vs.
time. After a time scale which is set by the non-linearity of the
spectrum, the average value of $N_{\mathbf{k}}$ and its standard
deviation are of similar size. However, the expectation value of
the number difference between mode $\mathbf{q}$ and mode
$\mathbf{k}-\mathbf{q}$ equals $M$ and is constant in time. The
standard deviation of that difference always remains zero, which
implies relative number squeezing between the two modes.

Thus far we have discussed the time evolution of the system within
the dressed state manifold, which is defined by the initial
population of the $\mathbf{k}$ and $\mathbf{q}$ modes, and have
neglected damping into empty modes. In analogy to the treatment of
spontaneous photon scattering as transfer between dressed states
manifolds of the dressed atom-laser system \cite{CCT}, we now
consider scattering into empty modes as transfer between dressed
state manifolds. Thus, the damping of excitations from mode
$\mathbf{k}$ is no longer elastic, but rather carries the energy
difference between the dressed states among which it occurred.

Evaluating $H_{B}=\frac{g}{2V}\sqrt{N_{0}}\sum_{\mathbf{q}'}
A_{kq'}\left(b_{\mathbf{q'}}^{+}b_{\mathbf{k}-\mathbf{q'}}^{+}b_{\mathbf{k}}\right)$,
between every pair of states in the two manifolds reveals the
spectral structure of the damping process.
We find that, the spectrum of the $N' =N-1$, $%
M' =N$ manifold, which has one less energy eigenvalue, is shifted
by roughly half the energy difference with respect to the $N=M$
energy spectrum. $H_{B}$ significantly couples only dressed states
with neighboring energies. This results in a structure of a
doublet in the Beliaev damping spectrum \cite{q damping}.

The finite lifetime of the dressed state results from the fact
that there is a quasi-continuum of $N' =N-1$, $M' =N$ manifolds,
all with an identical energy spectrum, to which a dressed state in
the $N=M$ manifold can couple to via $H_{B}$. To determine the
width of each transition between two dressed states, we use the
Fermi golden rule. The damping rate between a state $\left|
j\right\rangle _{N,M}$ in the $N$, $M$ manifold and a state
$\left| i\right\rangle _{N-1,M}$ in the $N-1$, $M$ manifold is
then given by

\begin{eqnarray}
&&\Gamma =\frac{2\pi}{\hbar}\sum _{\mathbf{q}'}
\frac{g^{2}N_{0}}{2V^{2}}\left| A_{kq'}\right| ^{2}\left|
_{N-1,M}\left\langle i\right|
b_{\mathbf{k}}\left| j\right\rangle _{N,M}\right| ^{2}\times \\
\nonumber &&\delta \left( \varepsilon _{\mathbf{k}}+\varepsilon
_{0}-\varepsilon _{\mathbf{q}'}-\varepsilon
_{\mathbf{k}-\mathbf{q}'}\right),
\end{eqnarray}

where $\varepsilon _{0}$ is the energy difference between $\left|
j\right\rangle _{N,M}$ and $\left| i\right\rangle _{N-1,M}$. Using
momentum conservation the energy conservation $\delta $-function
becomes a geometrical condition on the angle $\theta $ between
$\mathbf{k}$ and $\mathbf{q}'$

\begin{eqnarray}
&&\cos \left( \theta \right) =\\ \nonumber &&\frac{1}{2kq^{\prime
}}\left[ k^{2}+q^{\prime 2}+1-\sqrt{\left( \varepsilon
_{k}+\varepsilon _{0}-\varepsilon _{q^{\prime }}\right)
^{2}+1}\right],
\end{eqnarray}

where momentum is in units of $\hbar/\xi $, and energy is in units
of $gn$. Equation (6) is then simplified to

\begin{eqnarray}
&&\Gamma =\frac{g}{2\xi^3h}\int \left| A_{kq'}\right| ^{2}\left|
_{N-1,M}\left\langle i\right|b_{\mathbf{k}}\left| j\right\rangle
_{N,M}\right| ^{2}\times \\ \nonumber
&&\frac{q^{\prime }}{2k}%
\frac{\left( \varepsilon _{k}+\varepsilon _{0}-\varepsilon
_{q^{\prime }}\right) }{\sqrt{\left( \varepsilon _{k}+\varepsilon
_{0}-\varepsilon _{q^{\prime }}\right) ^{2}+1}}dq^{\prime }.
\end{eqnarray}

The spectrum of each transition is taken as a Lorenzian with a
width of $\Gamma/2\pi N_{\mathbf{k}}$ and a normalization of
$\Gamma/N_{\mathbf{k}}$ around $\epsilon_0$. $N_{\mathbf{k}}$ is
the average occupation of mode $\mathbf{k}$ for the two dressed
states involved. Averaging over all of the possible transitions
between the manifolds, Fig. 4 shows the damping spectrum between
the $N=M=5\times10^3$ and the $N=5\times10^3-1$, $M=5\times10^3$
manifolds, for the same model system as in Fig. 3 for $k$=3.2,
1.6, and 0.7 $\hbar/\xi$, and $q=k/\sqrt2$ \cite{scaling}. A clear
doublet structure is evident. The inset of Fig. 3 shows the
energy-conserving surfaces for the two center energies of the
$k$=0.7 $\hbar/\xi$ curve (solid line), and the energy conserving
surface for elastic damping from the same mode (dashed line).

\begin{figure}[h]
\begin{center}
\mbox{\psfig{figure=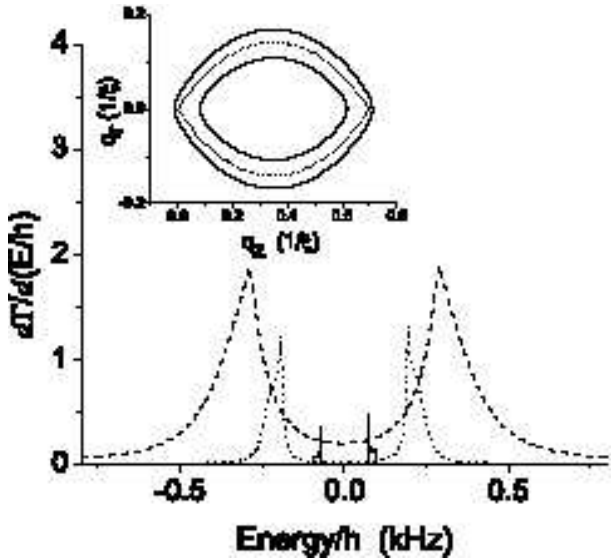,width=8.0cm}}
\end{center}
\vspace{0.4cm} \caption{Damping spectrum between the
$N=M=5\times10^3$ manifold and the $N=5\times10^3-1$,
$M=5\times10^3$ manifold, for the same condensate as in Fig. 2.
The three curves are for $k$=3.2 (dashed), $k$=1.6 (dotted) and
$k$=0.7 $\hbar/\xi$ (solid line), $q=k/\sqrt{2}$. The inset shows
the energy-conserving surfaces for the two center frequencies of
the $k=0.7$ $h/\xi$ curve (solid line) and the energy conserving
surface for elastic damping from mode $\mathbf{k}$ (dashed line).}
\end{figure}

The separation between the peaks in Fig. 4, which is equal to the
oscillation frequency, decreases with $k$ as $A_{kq}$. The
decrease in the width of each peak as a function of $k$ comes from
two contributions. Firstly, since damping into empty modes is
incoherent, it decreases as $\left| A_{kq'}\right| ^{2}$.
Secondly, as $k$ decreases there are less allowed empty modes on
the energy-conserving surface. Therefore, the doublet structure is
more resolved for the lower $k$ values. The difference in width
between the positive and negative energy peaks is mainly due to
the difference in the number of allowed empty modes on the energy
conserving surfaces. For small enough $k$ values, with respect to
the oscillation frequency, the energy conserving surface for the
negative energy peak completely disappears and only the positive
energy peak remains. Another contribution to the width of the
resonance in Fig. 4 is due to the non-linearity of the dressed
state spectrum, which gives a slightly different energy for
transitions between different dressed state pairs.

Several other mechanisms contribute to the broadening of the two
peaks which are not included in Fig. 4. The fact that only the
first scattering event occurs between the $N=M=5\times10^3$ and
the $N=5\times10^3-1$, $M=5\times10^3$ manifolds will further
broaden the resonances. Since the energy splitting scales as
$\sqrt{N}$, in an experiment where one scatters $dN$ atoms from
mode $\mathbf{k}$, this will result in a relative broadening of
~$\frac{dN}{2N}$. According to the same scaling, an initial
coherent, rather than Fock, state will cause a relative broadening
of $\frac{\sqrt{N}}{2N}$. The condensate finite size or
inhomogeneous density profile, will further contribute to the
width of the resonance. We estimate that for the experimental
parameters of \cite{our experiment1} a doublet structure in the
Beliaev damping spectrum can be resolved. Experimentally, the
energy doublet can be observed by computerized tomography analysis
of time of flight absorption images of the 3WM system \cite{our
experiment2}.

Another process which will transfer the system between manifolds
is that of a two-photon Bragg transition from the condensate to
either the $\mathbf{k}$ or $\mathbf{q}$ modes \cite{our
experiment1}. The line-shape of the Bragg transition into the
$\mathbf{k}$ mode also splits into a doublet structure \cite{tbp}.

The dressed state formalism can be applied to other processes
which involve bosonic amplification. We have applied this method
to a strongly Bragg-driven condensate, with large momentum
transfer, by diagonalizing $H_{int}=\frac{h\Omega _{R}}{2}\left(
a_{\mathbf{k}}^{+}a_{0}+a_{0}^{+}a_{\mathbf{k}}\right)$, $\Omega
_{R}$ is the two-photon Rabi frequency, and to atomic 4WM with
large relative momentum by diagonalizing
$H_{int}={g\over{2V}}\left(
a_{\mathbf{k}}^{+}a_{0}^{+}a_{\mathbf{k}-\mathbf{q}}a_{\mathbf{q}}+a_{\mathbf{k}-\mathbf{q}}^{+}a_{\mathbf{q}}^{+}a_{\mathbf{k}}a_{0}\right)$.
Both systems show a doublet structure in the spectrum of
collisions from a certain momentum mode, due to rapid oscillations
in the population of that mode \cite{tbp}.

In conclusion, we calculate the wave-mixing dynamics between
three, low $k$, Bogoliubov quasi-particles in a Bose condensate.
The Hamiltonian of this system is diagonalized to the next order
in $\sqrt{N_0}$. The resulting basis of dressed states allows for
the efficient, linear, propagation of the system in time.
Non-linear oscillations between the different momentum modes are
observed. Relative number squeezing between the $\mathbf{q}$ and
the $\mathbf{k}-\mathbf{q}$ momentum modes is shown. Beliaev
damping of excitations from these modes is treated as a transfer
between dressed states manifolds. The damping process is shown to
become inelastic and, similarly to the optical Mollow triplet,
exhibits a doublet-like energy spectrum, which is more resolved
for the low $k$ values.

We thank Pierre Meystre and Eitan Rowen for helpful discussions.
This work was supported by the Israeli Science Foundation.


\begin{references}
\bibitem{Phillips fwm}  L. Deng et. al., Nature {\bf 398}, 218 (1999).
\bibitem{Superadiance exp}   S. Inouye et., al. Science {\bf 285}, 571 (1999).
\bibitem{Amplification Kett} S. Inouye et. al., Nature {\bf 402}, 641 (1999).
\bibitem{Amplification Japan}  M. Kozuma et. al., Science {\bf 286}, 2309 (1999).
\bibitem{Julienne1}  M. Trippenbach, Y. B. Band and P. S. Julienne, Phys. Rev. A {\bf 62},
023608 (2000).
\bibitem{Myestre1}  E. V. Goldstein and P. Meystre, Phys.
Rev. A {\bf 59}, 3896 (1999).
\bibitem{Vogels} J. M. Vogels et. al., Phys. Rev. Lett. 88, 060402 (2002)  .
\bibitem{Bogoliubov}  N. N. Bogoliubov, J. Phys. (USSR) {\bf 11}, 23 (1947).
\bibitem{Burnett1}  S. Morgan, S. Choi, K. Burnett, and M. Edwards, Phys. Rev. A {\bf 57},
3818 (1998).
\bibitem{CCT}  C. Cohen-Tannoudji, J. Dupont-Roc and G. Grynberg, {\it Atom-Photon Interactions} (John Wiley and Sons inc.,
1998), ch. VI.
\bibitem{Burnett2}  J. Rogel-Salazar, G. H. C. New, S. Choi and K. Burnett,  Phys. Rev. A {\bf 65},
023601 (2002).
\bibitem{Foot1}  E. Hodby, O. M. Marago, G. Hechenblaikner and C. J. Foot,  Phys. Rev. Lett. {\bf 86},
2196 (2001).
\bibitem{Fetter}  A. Fetter. in { \it Proceedings of the International School of Physics 'Enrico Fermi', Course CXL } (IOS Press,
1999).
\bibitem{our experiment1}  J. Steinhauer, R. Ozeri, N. Katz and N. Davidson, Phys. Rev. Lett. {\bf 88}, 120407 (2002).
\bibitem{Pitaevskii}  L. Pitaevskii and S. Stringari, Phys. Lett. A {\bf 235},
398 (1997).
\bibitem{Giorgini}  S. Giorgini, Phys. Rev. A {\bf 57},
2949 (1998).
\bibitem{our experiment3}  N. Katz, J. Steinhauer, R. Ozeri and N. Davidson,  Phys. Rev. Lett. {\bf 89}, 220401 (2002).
\bibitem{scaling law} This scaling law was verified numerically over several $N=M$ values. A similar scaling law can be written for the $N\neq{M}$ case.
\bibitem{Fock}  Since excitations
will usually bear an uncertainty of $\sqrt{N}$ in their
population, the choice of an excitation Fock state as the system
initial condition is somewhat unrealistic. This will smear the
system over an ensemble of manifolds. However, for large $N,M$ the
relative spread in the excitation population becomes less and less
significant.
\bibitem{q damping}  Damping from the $\mathbf{q}$ momentum mode can be treated in a similar manner. Since the population in the $\mathbf{q}$ mode does not reach 0, the damping spectrum has the structure of a triplet.
\bibitem{scaling}  Since we cannot diagonalize matrices of size $5\times10^3$ by $5\times10^3$, we diagonalize smaller $N$,$M$ matrices and scale the resulting energies by a factor of $\sqrt{N}$.
\bibitem{our experiment2}  R. Ozeri , J. Steinhauer, N. Katz and N. Davidson, Phys. Rev. Lett. {\bf 88}, 220401 (2002).
\bibitem{tbp}  R. Ozeri et. al., (To be Published).
\end{references}
\end{document}